\documentclass[a4paper,12pt]{article}
\usepackage[cp1251]{inputenc}
\usepackage[dvips]{graphicx}
\graphicspath{{.}}

\begin{document}
\mathsurround=2pt \sloppy
%\begin{center}
\title{\bf Influence of correlated impurities on the onset of Cooper pairing}
\author{I. A. Fomin\\
{\it P. L. Kapitza Institute for Physical Problems},\\{\it
Kosygina 2,
 119334 Moscow, Russia}}
%\end{center}
\maketitle
\begin{abstract}
Effect of two-particle correlations between impurities on the
temperature of transition of a Fermi liquid in the superfluid or
superconducting state is analyzed. It is shown, that correlations
with a radius, exceeding correlation length of the superconductor
can have a pronounced effect on the transition temperature. The
equation for the transition temperature is corrected for
correlations of the impurities. Possible applications of the new
equation are discussed.

\end{abstract}
{\bf 1.} Effect of impurities or vacancies on thermodynamic
properties of traditional superconductors is well described by the
Abrikosov and Gorkov (AG in what follows) theory of
superconducting alloys \cite{AG}. Possible correlations among the
impurities are neglected in this theory. Difference with respect
to the pure superconductor in this theory is characterized by one
parameter - electron mean free path $l$ relative to the coherence
length of the superconductor $\xi_0$. It has been observed
recently \cite{nature} (and references therein), that certain
impurities in the high-T$_c$ compounds demonstrate a tendency to
short-range ordering, which influences the superconducting
transition temperature T$_c$ of the compound. Even more clear
example of such influence is presented by the superfluid $^3$He in
a high-porosity aerogel \cite{parp1,halp}.  Aerogel is a rigid
structure consisting of strands of a diameter $a\approx$3 nm  and
the average distance between the strands $\xi_a\approx$40 nm for
98\% porosity. The latter distance is usually taken as a
characteristics of correlations within aerogel.
 Thermodynamic properties of $^3$He in aerogel, in particular
dependence of the suppression of the transition temperature with
respect to that of pure $^3$He on $\xi_0/l$ demonstrate
significant deviation from the universal behavior, predicted by
AG-theory. This deviation was attributed to the existence of
structural correlations in aerogel with a characteristic radius
which is comparable with $\xi_0$ \cite{parp2}. In 98\% aerogel
depending on a pressure $\xi_0$ for $^3$He varies from infinity at
P$\approx$5 bar to $\xi_0\approx$20 nm at solidification. The
condition $\xi_a\approx\xi_0$ is met at P$\approx$15 bar.

 In the present paper effect of correlations on T$_c$ is
considered under the assumption that the correlations are weak.
Parameter, which characterizes deviation from the AG-theory turns
out to be the ratio $\frac{R^2}{\xi_0l_{tr}}$, where $R$ is a
length, characterizing correlation of impurities. For aerogel
$R\sim\xi_a$ and the effect of correlations on the T$_c$ can be
significant. Analogous approach was used before for the situation
when suppression of T$_c$ is small in comparison with the original
T$_c$ \cite{fomin}. Here this restriction is lifted and the
argument is extended for arbitrary suppression of T$_c$. This
generalization is necessary for $^3$He in aerogel since at a
pressure about 6 $bar$ the observed T$_c$ is zero so that the
suppression of T$_c$ is equal to T$_c$ itself.

{\bf 2.}  Transition temperature T$_c$ is defined as a temperature
of the onset of the long-range order. At this temperature appears
a pole in the thermodynamic vertex part $\Gamma(\Omega_n,{\bf q})$
for scattering of two quasi-particles with the total momentum
 ${\bf q}=0$ and the total frequency $\Omega_n=0$ \cite{AGD}.
Equation for T$_c$ has the standard form for all types of Cooper
pairing:
$$
\frac{1}{\lambda}=\Pi(0,0), \eqno(1)
$$
where $\lambda$ is a constant of the pairing interaction and
polarization operator is defined as
$$
\Pi(\textbf{q},\Omega_n)=T\sum_m\int\frac{d^3k}{(2\pi)^3}
G(\textbf{k+q},\omega_m+\Omega_n)G(\textbf{-k},\omega_{-m})\equiv
T\sum_m L^{(0)}_m(\textbf{q},\Omega_n),   \eqno(2)
$$
$G(\textbf{k},\omega_{n})$ etc. are thermodynamic Green functions
of quasi-particles.  In what follows
$L^{(0)}_m(\textbf{q},\Omega_n)$ always enters at
 $\Omega_n=0$, and a shorthand notation
$L^{(0)}_m(\textbf{q},0)\equiv t_m(\textbf{q})$ is used.

Effect of impurities is described by the potential
$$
U({\bf r})=\sum_a u({\bf r-r}_a)         \eqno(3)
$$
which is a sum of individual potentials of impurities, situated in
 positions ${\bf r}_a$. These positions are random.  In the
standard technique \cite{AGD} Green functions of quasiparticles
are expanded in the Born series over $U({\bf r})$ and averaged
over positions of the impurities. The sum of the obtained series
can be written as:
$$
\langle G(\varepsilon;{\bf k},{\bf k'})\rangle=2\pi\delta({\bf
k}-{\bf k'}) [G_0^{-1}-\langle\Sigma\rangle]^{-1},     \eqno(4)
$$
where $\langle\Sigma\rangle$ is the averaged self-energy.
  Two first terms of the Born series for $\langle\Sigma\rangle$
are represented by the diagrams Fig. 1. Correlations among the
impurities introduce changes in the procedure of averaging.

\begin{figure}
\begin{center}
\includegraphics[width=0.75\linewidth,keepaspectratio]{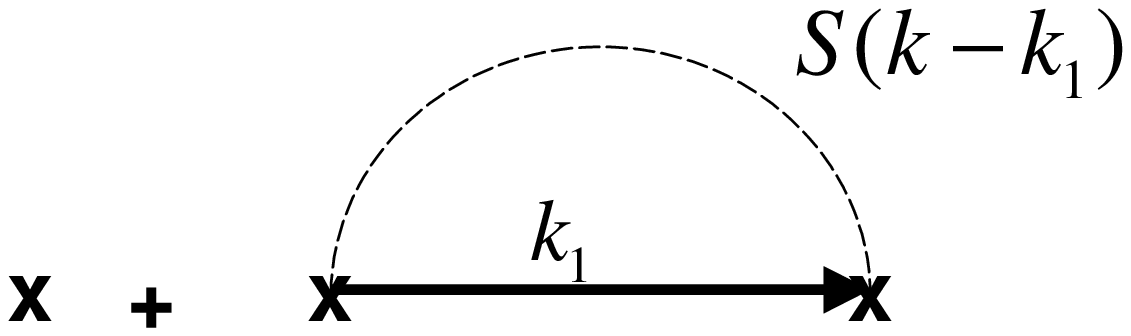}
\end{center}
\caption ~
\end{figure}

To every cross corresponds a function $u({\bf k'}-{\bf k})\sum_a
e^{i({\bf k}-{\bf k'}) {\bf r}_a}$, where ${\bf k}$  and ${\bf
k'}$  -- the in-coming and out-going momenta.  Essential
contribution comes from the term with two crosses. It is
proportional to the averaged sum: $\langle\sum_{a,b}e^{i({\bf
k}-{\bf k}_1){\bf r}_a+i({\bf k}_1-{\bf k'}){\bf r}_b}\rangle$.
Summation over ${\bf r}_a$ renders a factor $(2\pi)^3n\delta({\bf
k}-{\bf k'})$. The remaining sum is a structure factor
$$
S({\bf q})\equiv \langle\sum_{b}e^{i{\bf q}({\bf r}_b-{\bf
r}_a)}\rangle,        \eqno(5)
$$
where ${\bf q}={\bf k}_1-{\bf k}$. For uncorrelated impurities
$S({\bf q})=1+(2\pi)^3n\delta({\bf q})$. The unity comes from the
summand with ${\bf r}_b={\bf r}_a$. The term with $\delta({\bf
q})$ can be dropped, because in further calculations it enters
with the factor $u(0)$, which is excluded by a shift of the ground
state energy. When correlations are present there appears an
additional term $\bar{S}({\bf q})$ which is directly related to
the correlation function in the coordinate representation.
Summation in Eq.(5) for ${\bf r}_b\neq{\bf r}_a$ can be changed
for integration with the probability $w({\bf r}_b|{\bf r}_a)$ to
find a particle in the point ${\bf r}_b$ if there is a particle in
the point ${\bf r}_a$. In the isotropic case $w({\bf r}_b|{\bf
r}_a)$ depends on a relative distance $r=|{\bf r}_b-{\bf r}_a|$.
At $r\rightarrow \infty$ correlations vanish and $w(r)$ tends to a
constant. Normalization of $w(r)$ is chosen so that this constant
is unity. A measure of correlations is $v(r)=w(r)-1$ (cf.
\cite{landau5}). With these notations
$$
\bar{S}({\bf q})=n\int v(r)e^{-i{\bf q}{\bf r}}d^3r. \eqno(6)
$$
  and:
$$
\Sigma_2(\varepsilon;{\bf k})=n\int u({\bf k}_1-{\bf k})u({\bf
k}-{\bf k}_1) [1+\bar{S}({\bf k}-{\bf k}_1)]G(\varepsilon;{\bf
k}_1)\frac{d^3k_1}{(2\pi)^3}.     \eqno(7)
$$
Integration of two crosses in one block $\Sigma_2$ is graphically
 represented by the dashed line, connecting the crosses (cf.
 Fig.1). To this line corresponds now the sum $1+\bar{S}({\bf
 q})$. The averaged Green function has the same form as for
 uncorrelated impurities
$$
\langle G(\omega_n;{\bf k},{\bf k'})\rangle=(2\pi)^3\delta({\bf k}-{\bf k'})
(i\omega_n-\mu+\frac{i}{2\tau} sgn \omega_n)^{-1},      \eqno(8)
$$
except that the average inverse time between the collisions
includes $\bar{S}({\bf q})$:
$$
\frac{1}{\tau}=2\pi
 n\nu_0\int|u(\theta)|^2[1+\bar{S}(\theta)]\frac{d\Omega}{4\pi}.
\eqno(9)
$$
Here $\theta$ is the angle between  ${\bf k}_1$ and ${\bf k}$.The
new term $\bar{S}(\theta)$ takes into account effect of
interference of de Broigle waves of quasiparticles scattered by
different impurities.  This effect can be formally included in the
definition of cross-section
$\overline{|u(\theta)|^2}=|u(\theta)|^2[1+\bar{S} (\theta)]$. It
 is determined by the properties of the normal phase and does not
depend on T$_c$. The effect of interference is usually small,
since the average
 distance between the impurities $a$ is large in comparison with
 the wave length of  quasiparticles, and $\bar{S}({\bf q})$
 vanishes at $|{\bf q}|>1/a$. Contribution of $\bar{S}({\bf q})$
 to the integrand in Eq. (7) is comparable with unity only at
 angles $\theta\sim 1/ak_F$. For small concentration of impurities
 $x$  the relative contribution of correlations is on the order of
 $x^{2/3}$.

 At the averaging of the polarization operator except for
substitution of the averaged Green functions in Eq.(2) it is
necessary to take into account the contribution  of "ladder"
diagrams of the type represented by the first diagram in the
r.h.s. of the equation Fig.3. The dashed lines bring in the factor
$1+\bar{S}({\bf k}-{\bf k}_1)$, which can also be included in the
cross-section $|\overline{u({\bf k}_1-{\bf k})}|^2$.
 For definiteness the explicit analysis here is made for the
p-wave pairing. That makes possible to apply the obtained results
to the superfluid $^3$He in aerogel.  The interaction leading to
the p-wave pairing is proportional to the scalar product of unit
vectors $\hat{\textbf{k}}, \hat{\textbf{k}}\prime$:
$V(\textbf{k},\textbf{k}\prime)=3V_1
(\hat{\textbf{k}}\cdot\hat{\textbf{k}}\prime)$. As the result the
loops on the left and the right ends of the ladder diagrams
acquire factors $\hat{\textbf{k}}$ and $\hat{\textbf{k}}\prime$
respectively. The cross-section of scattering on the spherically
symmetric impurities can be expanded in a series of Legendre
polynomials:
$$
\overline{|u({\bf k}_1-{\bf
k})|^2}=\sigma_0+3\sigma_1(\hat{\textbf{k}}
\cdot\hat{\textbf{k}}\prime))+
5\sigma_2P_2(\hat{\textbf{k}}\cdot\hat{\textbf{k}}\prime))+... .
\eqno(10)
$$
After the angular integration in every loop of the ladder diagrams
only $\sigma_1$-component remains. It enters results via inverse
collision time $1/\tau_1=2\pi\nu_0\sigma_1$.   Summation of the
series for $\Pi(0,0)$ renders familiar equation for T$_c$ at the
p-wave Cooper pairing:
$$
1=2\pi\nu_0V_1\sum_{n\geq 0}\frac{1}{\omega_n+1/2\tau_{tr}},
\eqno(11)
$$
where $1/\tau_{tr}=2\pi n\nu_0(\sigma_0-\sigma_1)$. For the
pairing with other angular momenta $l\neq 1$ a proper $\sigma_l$
and $1/\tau_l$ have to be used. In particular for the case of
s-wave pairing the resulting $1/\tau$ would be zero and transition
temperature is not lowered in agreement with Refs. \cite{AG,andsn}

So, within the scheme of the AG theory, correlations among the
impurities enter the equation for T$_c$ only implicitly, via the
mean free path, determined by the normal phase.

{\bf 3.} A non-trivial effect of correlations occurs when the
contribution of diagrams with four crosses is taken into account
in $\Pi(\textbf{q},\Omega_n)$. Of particular interest for the
present discussion are the diagrams with three dashed lines as
shown on the Fig.2.

\begin{figure}
\begin{center}
\includegraphics[width=0.75\linewidth,keepaspectratio]{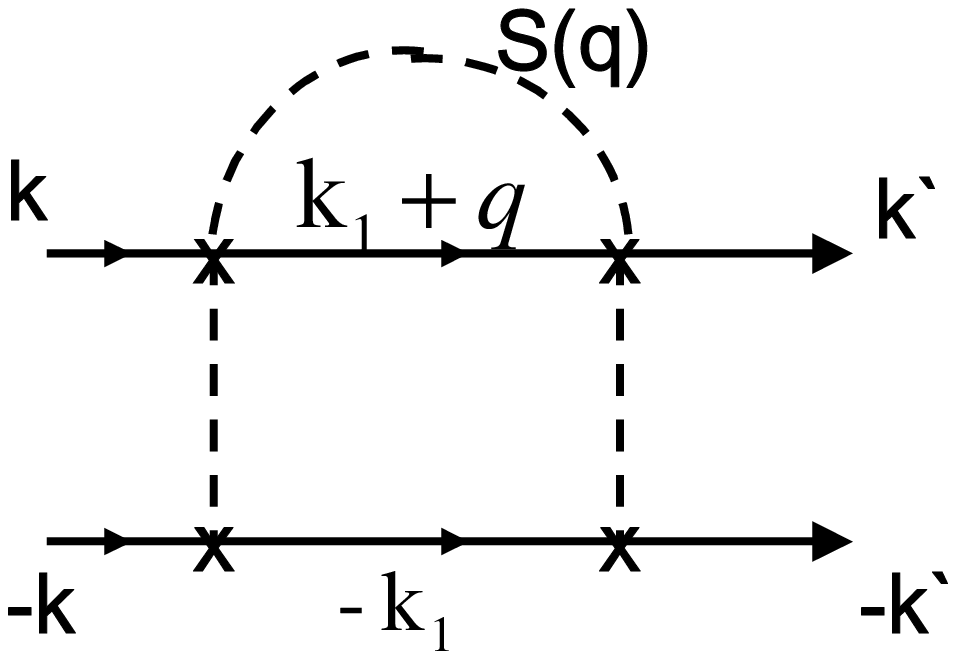}
\end{center}
\caption ~
\end{figure}

Without correlations it would correspond to the fourth order
correction to the amplitude of scattering of two quasiparticles
with zero total momentum on the same impurity. For
 interacting quasiparticles  in the intermediate state instead of
the product of two single particle Green functions $G_n({\bf
k}_1+{\bf q}_1) G_{-n}(-{\bf k}_1)$ the two-particle Green
function $G^{II}({\bf k}_1+{\bf q}_1,-{\bf
 k}_1,\omega_l,-\omega_l;{\bf k}_2+ {\bf q}_1,-{\bf
 k}_2,\omega_m,-\omega_m)$ has to be used.  As a function of ${\bf
 q}_1$ $G^{II}$ is singular at ${\bf q}_1=0$. This singularity
 corresponds to formation of a virtual Cooper pair. It increases
 the weight of states with ${\bf q}_1\sim 1/\xi_0$ in the
 integral, corresponding to this diagram. If the impurities are
 correlated on a distance $\sim \xi_0$ there remains a net effect
 of interference of the waves with ${\bf q}_1\sim 1/\xi_0$
 scattered by impurities. Depending on a sign of correlations it
 can decrease or increase the destructive effect of impurities.
 The aim of the following discussion is to find out when the
 effect of interference is essential.

 Interaction of electrons conserves the total momentum ${\bf q}_1$
 and the total frequency $\Omega=0$ but, unlike the elastic
 scattering by impurities, it does not conserve each of the
 frequencies $\omega_l$. As the result the polarization operator
 became a sum over two indices: $\Pi(0)=T^2\sum_{lm}L_{lm}(0)$,
 where $L_{lm}(0)$ is a block, represented by the black rectangle
 in the Fig.3 .

\begin{figure}
\begin{center}
\includegraphics[width=0.75\linewidth,keepaspectratio]{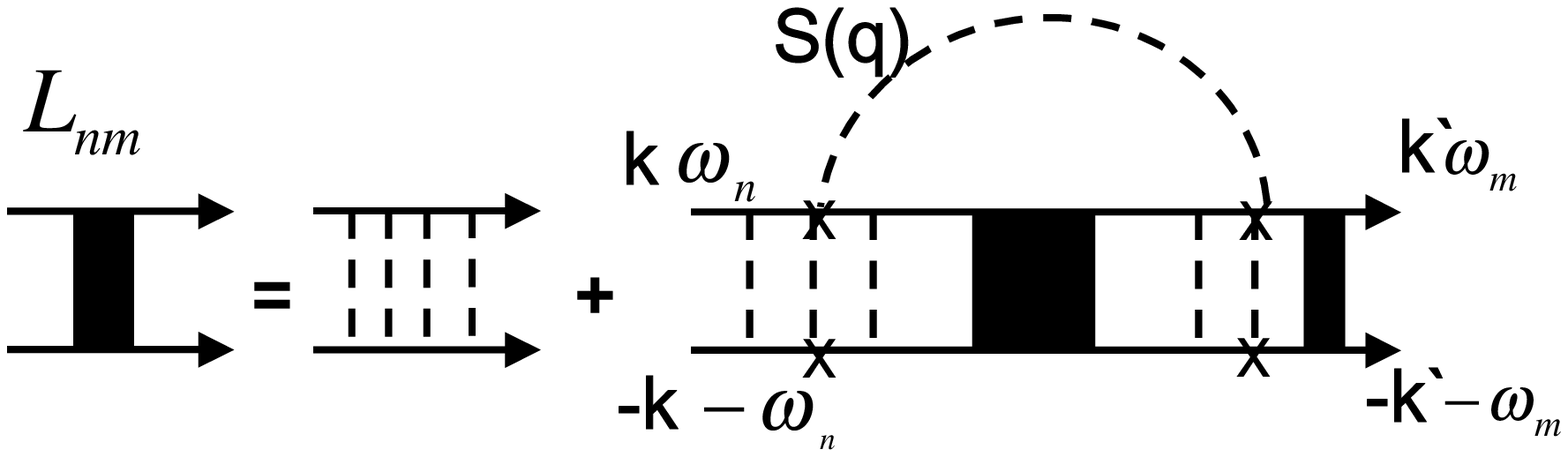}
\end{center}
\caption ~
\end{figure}

In analytic form the equation Fig.3 can be written as:
$$
\bar{L}_m=t_m(0)-t_m(0)T\sum_jM_{mj}\bar{L}_j, \eqno(12)
$$
where
$$
M_{mj}=n\sigma_1^2\int t_m({\bf q})\Gamma({\bf q})[1+
 \bar{S}({\bf q})]t_j({\bf q})\frac{d^3q}{(2\pi)^3},   \eqno(13)
$$
 $\bar{L}_j=T\sum_m L_{jm}(0)$ and  $t_j({\bf q})=B_j({\bf
q})/[1-n\sigma_1B_j({\bf q})]$ with $B_j({\bf
q})=\frac{2\pi\nu_0}{vq}
arctg\left(\frac{vq}{2|\tilde{\omega}_j|}\right)$. In a limit
$q\rightarrow 0$ it tends to $t_j(0)=2\pi\nu_0/|\tilde{\omega}_j|$
with $\tilde{\omega}_j=\omega_j+1/2\tau_{tr}$.

The matrix $M_{mj}$ is proportional to $\sigma_1^2$, which brings
in a small factor  $(\xi_0/l_{tr})^2$. A special situation arises
when the impurities are correlated on a distance $R$ which is
greater than $\xi_0$. In this case $\bar{S}({\bf q})$ is enhanced
in a region $q\sim 1/R$. In the limiting case $R\gg\xi_0$ all
functions under the integral in Eq.(13) except for $\bar{S}({\bf
q})$ and $\Gamma({\bf q})$ can be taken at $q=0$ so that
$M_{mj}=n\sigma_1^2t_m(0) Q t_j(0)$, where $Q=\int\Gamma({\bf
q})\bar{S}({\bf q})\frac{d^3q}{(2\pi)^3}$. At the transition
temperature and at small $q$ $\Gamma({\bf
q})=-\gamma/\nu_0(\xi_0q)^2$ with coefficient
$\gamma=12/7\zeta(3)$. A singular behavior of $\Gamma({\bf q})$ at
$q\to 0$ increases the weight of small $q$ in the integral and
since $\bar{S}({\bf q})$ is enhanced at $q\sim 1/R$ the value of
the integral is also enhanced. For estimation of integrals,
containing $\bar{S}$ the form $\bar{S}({\bf q})=A\delta(q-q_0)$
with $q_0=1/R$ is convenient. The coefficient $A$ is determined
from the normalization condition $\int\bar{S}({\bf
q})d^3q=(2\pi)^3nv(0)$, so that $A=2\pi^2R^2nv(0)$. Omitting
insignificant coefficients on the order of unity we arrive at
$$
Q\sim-\frac{R^2}{\xi_0^2}\frac{nv(0)}{\nu_0(2\pi)^2}.
\eqno(14)
$$
This estimation shows, that the second term in the rhs of Eq.(12)
relative to the first is on the order of $(R/l)^2$. At $R\ll l$
Eq.(12) can be solved by iterations. As a first iteration we have:
$$
\bar{L}_m=t_m(0)-t_m(0)T\sum_jM_{mj}t_j(0), \eqno(15)
$$
Summation of this equation over m with account of Eq.(1) after
standard transformations renders the equation for for $T_c$:
$$
\ln\frac{T_c}{T_{AG}}=\psi\left(\frac{1}{2}+\frac{1}{4\pi
T_{AG}\tau_{tr}}\right)-\psi\left(\frac{1}{2}+\frac{1}{4\pi
T_c\tau_{tr}}\right)-f(T_c), \eqno(16)
$$
where
$$
f(T_a)=n\frac{\sigma_1^2}{\nu_0}\int\left[T\sum_mt_m(0)t_m({\bf
q})\right]^2\Gamma({\bf q})[1+\bar{S}({\bf
q})]\frac{d^3q}{(2\pi)^3}.   \eqno(17)
$$
and $\psi(z)=\frac{d}{dz}\ln\Gamma(z)$ is di-gamma function.
Equations (16) and (17) form a closed system, which for given
material parameters and a given form of the structure factor
determines the transition temperature in a liquid with correlated
impurities. Difference with respect to the standard result
$T_{AG}$ is represented by the function $f(T_c)$ in the r.h.s. of
Eq.(16).

Further iterations of Eq. (12) render consecutive terms in the
expansion of the solution of Eq. (12) in powers of the parameter
$(R/l)^2$. The accuracy of Eq. (12) itself is limited by a
different parameter. The kernel (13) takes into account
interaction of the virtual Cooper pair only with two correlated
impurities. Averaging of diagrams including interaction with three
correlated impurities requires knowledge of the three-particle
correlation function. For estimation of the contribution of this
process we can decouple the three particle correlation function in
a product of the two-particle correlation functions and keep only
the most singular terms. The correction would contain extra factor
$Q\xi_0/l_{tr}\sim R^2/\xi_0l_{tr}$. Higher order correlations
 can be neglected if this parameter is small.

For unconventional superconductors the actual situation is
$\xi_0<l_{tr}$. We assume that the strong inequality $\xi_0\ll
l_{tr}$ is met. Then there exist a window $\xi_0\ll R\ll l_{tr}$.
 In a limit $R\gg\xi_0$ the asymptotic expression for $f(T_c)$ at
$(R/\xi_0)\rightarrow\infty$ can be used. The function $t_m({\bf
q})$ in the integrand of Eq. (17) can be taken at ${\bf q}=0$ and
the sum over m can be done explicitly
$$
T\sum_m
t_m^2(0)=\frac{\nu_0^2}{2T}\psi'\left(\frac{1}{2}+\frac{1}{4\pi
T_c\tau_{tr}}\right)     \eqno(18)
$$
The remaining integral over ${\bf q}$ in the limit $R\gg\xi_0$
 is proportional to $(R/2l_1)^2$ with a coefficient $\alpha\sim 1$
depending on the explicit form of the structure factor. The
resulting form of $f(T_c)$ is then:
$$
f(T_c)=-\alpha\left[\frac{R}{2l_1}\psi'\left(\frac{1}{2}+\frac{1}{4\pi
T_c\tau_{tr}}\right)\right]^2.   \eqno(19)
$$
The sign of $\alpha$ depends on a limiting value of the
correlation function $v(r)$ at $r\rightarrow 0$. At $v(0)>0$
$\alpha<0$, i.e. a tendency of impurities to form groups favors
formation of Cooper pairs and increases T$_c$ and vice versa.

{\bf 4.} For application of the obtained formulae to aerogel we
have to substitute its structure factor in the definition of $Q$.
 98\% Aerogel within the interval of distances $a\ll r\ll R$ has a
fractal structure with the dimensionality $D=1.6\div 1.8$
\cite{parp2}. Its structure factor within $1/R\ll q\ll 1/a$ varies
as $q^{-D}$. As a consequence the integral $Q\sim\int dq/q^D$ is
dominated by the lower limit $\sim 1/R$. Unfortunately, the
available data for the structure factor are given in arbitrary
units so that the value of $Q$ can be found only up to unknown
factor. Instead of the measured structure factor a model
expression was suggested \cite{fomin}, assuming that at $r>R$
correlations decay as $\exp(-r/R)$,  $R$ in this model is a
fitting parameter. Simplified analysis, using the limiting
expression valid for $R\gg\xi_0$ can be used only at pressures
above 20 bar. In this region Eqns. (16), (19) reproduce
qualitative features of the experimentally observed dependence of
$T_c$ on pressure. In particular, the difference $T_c-T_{AG}$ is
practically constant within this interval, but the interval itself
is not large enough for making definitive statements. In a region
P$<$20 bar the limiting expression for $t_m(0)$ is not
sufficiently accurate and Eqns. (16),(17) have to be solved with
the full $t_m({\bf q})$. That requires rather involved numerical
calculations which are not done yet.

 In conclusion, it is shown that effect of correlations on the
suppression of the transition temperature of a Fermi liquid in the
superfluid (superconducting) state by impurities can be essential
if correlation radius $R$ of impurities is greater than the
correlation length of a superconductor $\xi_0$. Correlations can
increase the transition temperature with respect to the value,
given by the standard theory of superconducting alloys.  The
obtained equations express the transition temperature in terms of
the structure factor of  ensemble of impurities. The quantitative
analysis takes into account only two-particle correlations. Within
this approximation a relative change of $T_c$ is characterized by
the parameter $R^2/\xi_0l_{tr}$. This parameter simultaneously
determines accuracy of the approximation. If $R^2/\xi_0l_{tr}\sim
1$, the results can be used only for qualitative estimations. Good
candidates for observation of the discussed effect are
superconductors with small values of $\xi_0$.

I am grateful to V.P.Mineev for pointing out to me the important
Ref. 2 and to I.M. Suslov for stimulating discussions. This
research was supported in part by RFBR grant 09-02-12131 ofi-m.


\begin{thebibliography}{99}




\bibitem{AG} A.A. Abrikosov and L.P. Gorkov, \emph{ZhETF}  {\bf 39}, 1781 (1961),
  [\emph{Sov. Phys. JETP} {\bf 12}, 1243 (1961)].

\bibitem{nature} M. Fratini, N. Poccia, A. Ricci, G. Campi, et.
   al., Nature, {\bf 466}, 841 (2010)

\bibitem{parp1} J. V. Porto and J. M. Parpia, {\it Phys. Rev. Lett.},
               {\bf 74}, 4667 (1995)
\bibitem{halp} D. T. Sprague,T. M. Haard,J. B. Kycia, V. R. Rand, Y. Lee,
P. Hamot and W. P. Halperin, {\it Phys. Rev. Lett.},
               {\bf 75}, 661 (1995)
\bibitem{parp2} J. V. Porto and J. M. Parpia, {\it Phys. Rev.},
               {\bf B59}, 14583 (1999).
\bibitem{fomin} I.A. Fomin, JETP Letters, {\bf 88}, 59 (2008)

\bibitem{AGD} A.A. Abrikosov, L.P. Gor`kov, I.E. Dzjaloshinskii,
Metody kvantovoi teorii polja v statisticheskoi fizike, Moskva
1962, Ch. VII.

\bibitem{landau5} L.D. Landau and  E.M. Lifshitz,  Statistical physics, part 1.
ch. 12--M.: Nauka, (1995).

\bibitem{andsn} P.W. Anderson, Journ. Phys. Chem. of Sol.,{\bf
11}, 26 (1959)






\end{thebibliography}
\end{document}